\documentclass{jjap3}  

\usepackage{graphicx}

\title{Relaxation of Cs atomic polarization at surface coatings characterized by X-ray photoelectron spectroscopy}

\author{Kotaro Kushida$^{1}$, Toshihiro Niwano$^{1}$, Takemasa Moriya$^{1}$, Tomohito Shimizu$^{1}$, Kazuki Meguro$^{2}$, Hideki Nakazawa$^{2}$, and Atsushi Hatakeyama$^{1*}$} 

\inst{$^{1}$Department of Applied Physics, Tokyo University of Agriculture and Technology, Koganei, Tokyo 184-8588, Japan  \\
$^{2}$Department of Electronics and Information Technology, Hirosaki University, Hirosaki, Aomori 036-8561, Japan\\
\rm E-mail: hatakeya@cc.tuat.ac.jp}

\abst{Paraffin coatings on glass slides were investigated through both X-ray photoelectron spectroscopy (XPS) and spin relaxation measurement for cesium (Cs) vapor.
The components of the glass substrate, such as silicon and oxygen, existed in the XPS spectra of the coated slides, indicating the imperfection of the prepared paraffin coatings.
The substrate was not observed after the annealing of the coatings in Cs vapor, which is known as a `ripening' process for spin relaxation measurement.
We found a general trend that effective anti-spin relaxation performance requires high paraffin and low Cs coverage on the surface. 
We also examined a type of diamond-like carbon film, anticipating the effect of anti-spin relaxation; our attempts have failed to date.}

\begin{document}
\maketitle

\section{Introduction}
\label{intro}
Alkali vapor cells are workhorses used in a variety of spin physics experiments with a combination of optical pumping techniques. The inner surfaces of the cells are often coated with some inert material that prevents spin relaxation of the polarized alkali atoms when they collide with the walls. These coatings, called spin anti-relaxation coatings, were first demonstrated in the late 1950s~\cite{Rob58}, and since then have been widely used in atomic physics experiments. The attraction of these coatings has recently renewed because they have been recognized as a key to many precision measurements, including optical magnetometry~\cite{Alx96,Bud98,Bud00,Cas09,Kim09} and atomic clocks~\cite{Rob82,Bud05,Guz06,Ban12}.
The most well-known anti-relaxation material for alkali vapors is paraffin, which can provide up to 10,000 bounces before atoms become depolarized~\cite{Bou66}.

Although anti-relaxation coatings have long been used, their functions are not fully understood. One example is the annealing process of `ripening'~\cite{Alx02,Sel10}, which enhances the anti-relaxation effect of the coatings~\cite{Gra05}. Another example is the recent dramatic improvement accomplished by using olefin~\cite{Bal10};  obtained one-minute spin relaxation times with this new coating suggests that there are some unknown key parameters and important mechanisms that contribute to the anti-relaxation effect.
From an application point of view, easy-to-fabricate and high heat-resistant coatings are required~\cite{Sel09}.

These facts have motivated a new class of experiments in place of traditional ones using sealed alkali vapor cells. An experimental system that used pairs of coated slides was constructed to measure spin relaxation times of alkali atoms colliding with the coating surfaces~\cite{Sel08}. Various standard surface analysis techniques were also introduced~\cite{Sel10}. These new studies revealed the surface morphology of the coatings\cite{Ram09,Hib12}, the indication of the importance of C=C double bonds~\cite{Sel10}, high-heat-resistant coating material~\cite{Sel09}, and so forth.
However, the systematic studies of each coating sample through both surface characterization and the spin relaxation mesurement are very limited\cite{Hib12}.

In this paper, we report the measurement of spin relaxation of cesium (Cs) atoms colliding with surface coatings characterized by X-ray photoelectron spectroscopy (XPS). 
We constructed an experimental system that enabled both spin relaxation measurement and surface analysis by XPS for each coating sample. Paraffin-coated glass slides were prepared in evacuated glass tubes and introduced into a relaxation measurement cell and an XPS chamber through a glove box, without exposure to air. 
This system helped us to understand the function of each process required for coated-cell preparation, including coating fabrication and ripening.

The components of the glass substrate, such as silicon (Si) and oxygen (O), were observed in XPS for prepared paraffin coatings, indicating that paraffin did not completely cover the glass slides. These glass components disappeared after the spin relaxation measurements, whereas Cs was observed. This modification can be attributed to the ripening process performed before the spin relaxation measurements. In terms of the relationship between coating performance and surface conditions, we observed a general trend, from about 30 pairs of coated slides, that effective anti-spin relaxation coatings required a high paraffin coverage, while having a small Cs composition on the surface. 
The O observed after the ripening process was found to bond to Cs.

We also conducted the experiment using a diamond-like-carbon (DLC) film as an anti-spin-relaxation coating.
We tested sp$^3$-rich DLC films. We did not observe any anti-relaxation effect, and we found that more Cs atoms were adsorbed onto the DLC film than on the paraffin coating. 

\section{Experiment}
\label{setup}

\begin{figure}[tbp]
 \begin{center}
  \includegraphics[width=80mm]{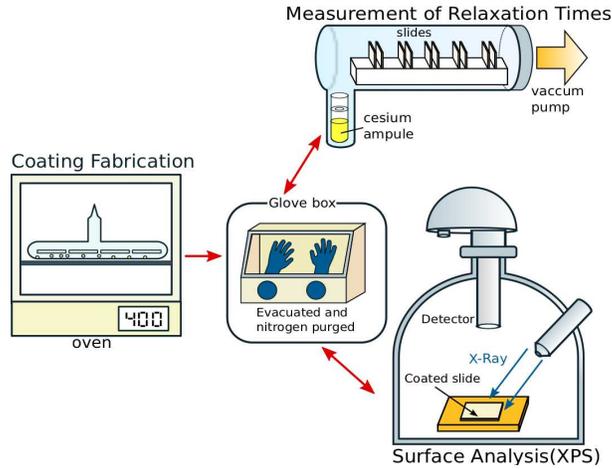}
 \end{center}
 \caption{Schematic of the experimental setup.}
 \label{Fig1}
\end{figure}

Figure~\ref{Fig1} shows the experimental system,
composed of four pieces of apparatus: the coating fabrication apparatus; the glove box; the spin relaxation measurement apparatus; and the surface analyzer equipped with XPS.
Using the glove box, we were able to transfer sample slides from one apparatus to another without exposing them to air.

\subsection{Coating preparation}
\label{coating}

We formed coatings on glass slides instead of on the inner walls of sealed cells, because it is easier to measure the surface conditions of the former. 
The slides were made of Pyrex, a type of borosilicate glass. The dimensions of the slides were $12.5\times25.0\times1.0$~mm$^3$. 

First, we describe the paraffin coating fabrication. Glass slides were washed by ultrasonic cleaning with ethanol and acetone. 
The paraffin used was Sasolwax H1;
its mean molecular weight is nominally 750~g/mol, corresponding to $n\sim50$ in $\rm{C_{n}H_{2n+2}}$, and its melting point is 112$^\circ$C. 
We did not distill it before use.

The coating fabrication process was similar to the ones employed for sealed cells~\cite{Cas09,Alx02}. Glass slides were set in a glass tube attached to a vacuum system with a paraffin reservoir. The tube was then evacuated and baked at 400$^\circ$C. The pressure reached a few $10^{-5}~$Pa. Pieces of the paraffin were introduced into the glass tube, which was then sealed. The paraffin coatings were formed on the slides by heating the glass tube at 400$^\circ$C for 4 h and then cooling it. The formed coatings had relatively rough surfaces and the typical thickness was a few $\mu$m.

The other material we evaluated was a DLC, an amorphous hydrogenated or non-hydrogenated form of carbon. It is composed of carbon (C) and hydrogen (H) as paraffin. The fabrication processes and the properties of various types of DLC films have been extensively studied. The composition ratio between C and H and the structure of DLC can be well controlled, and we therefore considered that it was a good candidate material to investigate key parameters for high antirelaxation effects. Note that DLC has been used in a basic research for storage of polarized ultracold neutron~\cite{Bry05} as well as in many commercial products.  

In this study we generated DLC coatings on Pyrex slides using the pulsed laser deposition (PLD) technique. 
A high-powered pulsed laser (KrF excimer laser, wavelength: 248~nm, pulse energy: 265~mJ/pulse, repetition rate: 20~Hz, pulse duration: 30~ns) irradiated a graphite (99.9\%) target under a  hydrogen atmosphere (1~Pa). After vaporization, C was deposited as a thin film on the Pyrex substrates at room temperature. DLC coatings formed by deposition for 90~s were 30~nm in thickness and had an sp$^3$-rich structure~\cite{Rob97} like diamond. It is thought that diamond may have anti-relaxation properties~\cite{Ste94}. 
The ability of DLC to withstand very high temperatures would be beneficial to applications requiring high alkali vapor densities.

It is noted that the paraffin coatings were not exposed to air before spin relaxation measurement and XPS analysis, while the DLC coatings were once exposed to air when they were taken out of the PLD chamber.

\subsection{Spin relaxation measurement}
\label{T1_1}

\begin{figure}[bt]
\resizebox{8cm}{!}{%

  \includegraphics{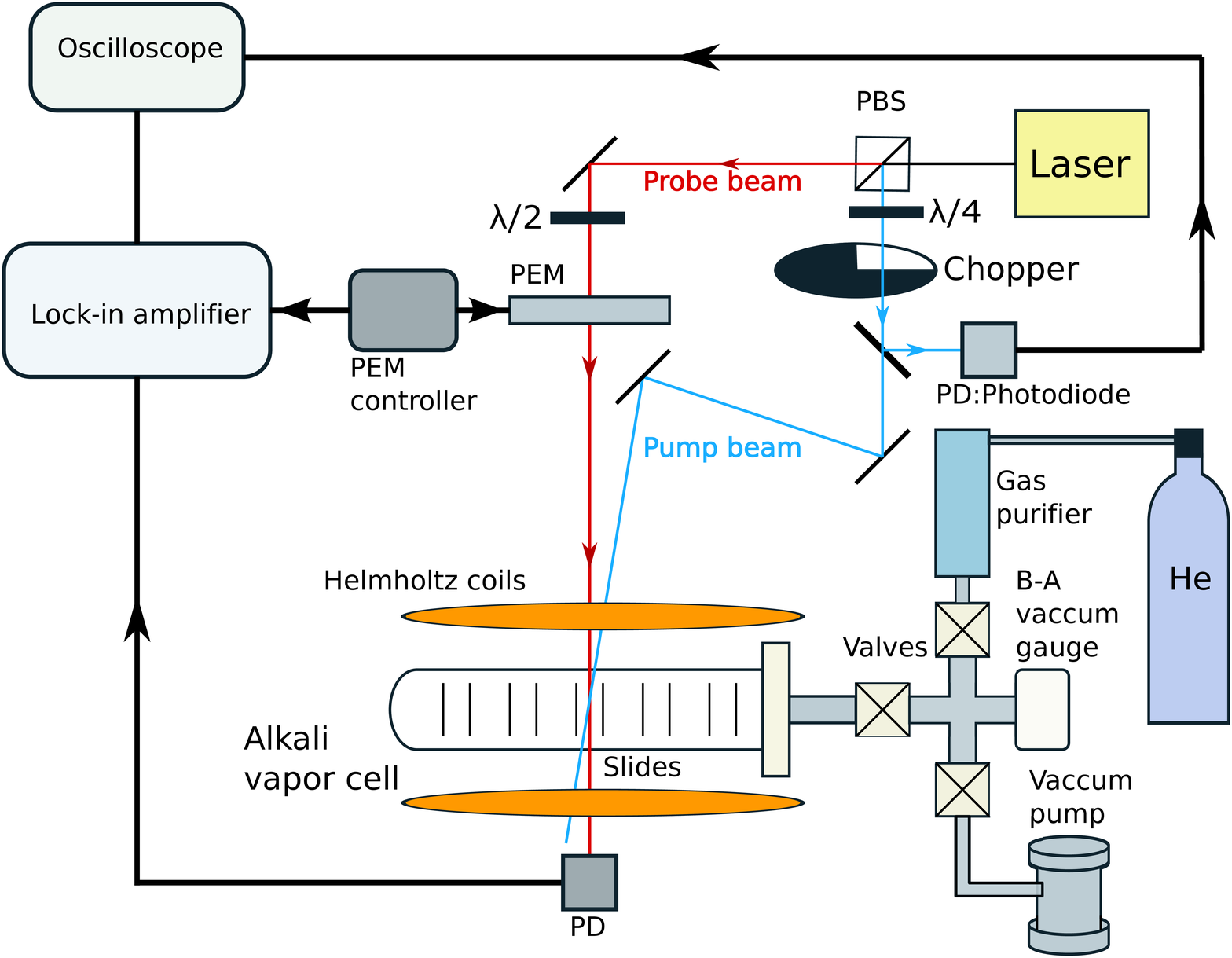}
}
\caption{Schematic of the setup for spin relaxation measurement.}
\label{Fig2}       
\end{figure}

The method of spin relaxation measurement was the same as the one reported in Ref.~\citen{Sel08}. 
The setup for the spin relaxation measurement is shown in Fig.~\ref{Fig2}. 
The main body of the alkali vapor cell was a Duran glass tube (inner diameter: 41.4 mm) with a NW40 flange. Coated glass slides were introduced to the cell in the glove box. They were set at equal intervals (2.0~mm) in the cell. We were able to evaluate up to five pairs of slides simultaeously. A Cs ampule was placed in a stem attached to the cell. 
Using dry scroll and turbomolecular pumps, the pressure of the cell reached a few $10^{-4}$~Pa. 

Gaseous Cs atoms were not initially observed in the cell. To obtain a high Cs vapor density, we performed ripening. We heated the main body of the cell to 80$^\circ$C  and 
the Cs ampule to 60$^\circ$C for 15 h.

When spin relaxation times were measured, helium (He) gas was introduced as a buffer gas, to prevent the Cs vapor escaping from between the slides. The He gas was of 99.99995\% purity and was used after passage through a gas purifier, which nominally reduced water,  carbon monoxide, carbon dioxide, and oxygen from 50 ppm to less than 1 ppb.
The main body of the cell, coated slides, and He gas were at room temperature, while the Cs ampule was heated to 50$^\circ$C. Spin relaxation times were measured as a function of He gas pressure from 10 to 60 Torr. 

The atomic spin polarization was produced and measured with a distributed feedback laser whose wavelength matched the Cs $D1$ transition between $F=4$ and $F'=3$ hyperfine levels. The laser beam was separated by a polarizing beam splitter (PBS) into pump and probe beams. The pump beam was circularly polarized with the quarter wavelength plate and entered between a pair of slides. It polarized the Cs atoms in the pump beam direction (the $z$ direction) by optical pumping. When the pump beam was shut off by the chopper, the probe beam detected the longitudinal relaxation of Cs atomic polarization. The probe beam was modulated between the $\sigma^{+}$ and $\sigma^{-}$ circular polarizations at 42 kHz using the photoelastic modulator (PEM). The difference in absorption between the $\sigma^{+}$ and $\sigma^{-}$ probe beams was detected by the lock-in amplifier. The measured signal, referred to as polarization signal in this paper, was proportional to the product of the spin $z$ component $\langle F_z\rangle$ of the $F=4$ hyperfine level and the $F=4$ population. The probe beam was weak enough not to affect relaxation signals.

A magnetic field ($2\times 10^{-4}$~T) was applied with a pair of Helmholtz coils along the direction of the two laser beams.

Obtained relaxation curves decayed exponentially with two time constants.  
We used the time constant of the faster exponential component to derive the number of bounces required for depolarizing Cs atoms as described in Ref.~\citen{Sel08} ; the slower component was supposed to originate from atoms outside a pair of slides.

\subsection{Surface analyzer and glove box}
\label{XPS_1}
We employed XPS, a widely used surface analysis technique, particularly for insulators. It is a non-destructive analysis method, 
and has been used in the studies of anti-relaxation coatings~\cite{Sel10}. XPS can observe elements within a few nm below the surface; the depth is determined by the inelastic mean free path of the photoelectron.

Our surface analysis apparatus was composed of an XPS chamber, a load lock chamber, and a glove box. 
The glove box was of vacuum type.  A 99.99995\% pure nitrogen gas filled the glove box after it was evacuated to about 100 Pa with a dry scroll pump. The dew point in the glove box was less than $-30^\circ$C, which corresponds to a moisture concentration of 400 ppm, or less, by volume. The load lock chamber, connecting the glove box and the XPS chamber, was evacuated using a turbomolecular pump after a coated slide was introduced from the glove box. The coated slide was then transferred to the XPS chamber. Further evacuation with an ion pump took the pressure in the XPS chamber to $1 \times10^{-7}$~Pa. 

An Al K$\alpha$ X-ray (1486.6 eV) was used to obtain the XPS spectra.
The observed area on a glass slide was 5$\times$10~mm$^2$.
The atomic concentration and the atomic environment, from peak areas and peak shifts in the spectra, respectively, were derived. 
We removed baselines from spectrum peaks using the Shirley method~\cite{Pro82} and fitted them with Voigt functions.
Observed peaks originated from Si, O, boron (B), sodium (Na), C, and Cs. Note that H is not detected by XPS. We derived the atomic concentration from peak areas normalized by ionization cross-sections and mean free paths. To adopt the values of mean free paths from the literature, we assumed a SiO$_2$ substrate~\cite{Tan88} for photoelectrons from the Pyrex glass components (Si, B, Na, and pre-ripening O) and a 26-n-paraffin substrate~\cite{Tan94} for those from C, Cs, and post-ripening O (see Sect. 3).
For the DLC coatings, we adopted the mean free paths in a glassy carbon substrate~\cite{Tan91}.
We assumed a uniform distribution of elements in substrates for all estimations of concentration~\cite{Hib13}.

To calibrate the binding energies of the spectra, we adjusted the Si 2p peak to 103.5 eV ($\rm{SiO_2}$)~\cite{Sha01} when Si peaks were observed.
This calibration resulted in a C 1s peak energy of 284.8 eV, which was used as a reference when Si was not observed;  we neglected a possible small shift of the C 1s peak due to exposure to alkali vapor~\cite{Hib13}.

\section{Results and discussion}
\label{results}
\subsection{Paraffin}
\label{Paraffin_1}

\begin{figure}[tb]
 \begin{center}
  \includegraphics[width=90mm]{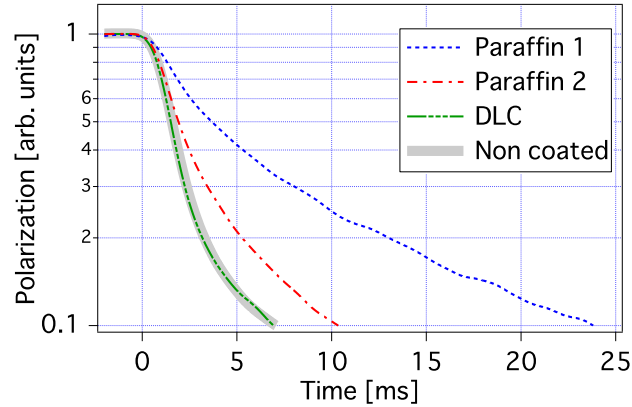}
 \end{center}
 \caption{Spin relaxation curves for four pairs of coated slides measured at a He gas pressure of 60 Torr. Two pairs of paraffin-coated slides, denoted as paraffin 1 and paraffin 2, were examined in a single measurement, while DLC- and non-coated slides were examined in a second measurement. The polarization curves were normalized at 0~ms, at which point the pump beam was shut off.}
 \label{Fig3}
\end{figure}

Typical spin relaxation data at a He gas pressure of 60 Torr are shown in Fig.~\ref{Fig3}. 
Two paraffin coatings, denoted as paraffin 1 and paraffin 2 in the figure, were fabricated under nominally identical conditions in the same batch and measured in the cell at the same time.
The probed polarization signals decayed after the pump beam was shut off at 0 ms. 
Spin relaxation for the paraffin coatings was slower than those for the non-coated glass slides. We therefore confirmed that the fabricated paraffin coatings had a spin anti-relaxation effect. 
The numbers of bounces for paraffin 1 and 2 were derived from the relaxation times of 2.5~ms and 1.9~ms to be 110 and 70, respectively. 
The uncertainty in evaluated bounces was approximately 10\%, originating from the He pressure, the substrate interval, and the diffusion constant~\cite{Ish99}.  
The numbers of bounces for non-coated glass slides were less than a lower bound of bounces (about 10) evaluable in this experiment. 

\begin{figure}[tb]
 \begin{center}
  \includegraphics[width=90mm]{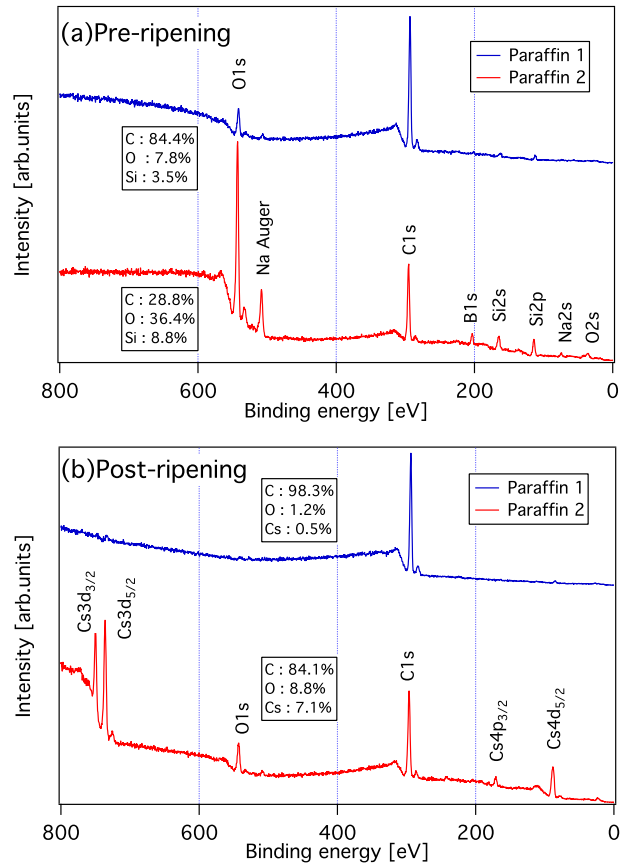}
 \end{center}
 
 \caption{XPS spectra taken (a) before and (b) after the measurement of relaxation time.  They are referred to as `pre-ripening' and `post-ripening', respectively. Atomic concentrations are shown for the main elements.} 
 \label{Fig4}
\end{figure}

Figure~\ref{Fig4} shows XPS spectra for paraffin 1 and paraffin 2 taken before and after the spin relaxation measurement.
In the pre-ripening spectra [Fig.~\ref{Fig4}(a)] obtained before the spin relaxation measurement, 
peaks originating from Si, O, Na, and B were observed for both paraffin coatings. They are components of the Pyrex glass substrates, indicating that the substrates were not covered perfectly before ripening. This difficulty in forming perfect coatings may be one of the causes of variations in anti-relaxation performance of paraffin-coated cells~\cite{Cas09,Sel10}.
 Paraffin 1, which has better anti-relaxation performance, shows relatively low Si peaks and a high C peak compared to paraffin 2. Because the C peak indicates the presence of paraffin on the surface, this result leads to the reasonable conclusion that high initial paraffin coverage is required for high spin anti-relaxation performance.

After the ripening process, 
the glass-originated Si, Na and B peaks disappeared, as shown in Fig.~\ref{Fig4}(b), while Cs peaks appeared. 
Furthermore, the C concentration in the post-ripening spectra increased compared to the pre-ripening spectra.
We therefore conclude that during the ripening process, paraffin (and Cs) covered the substrates more effectively. This increase in paraffin coverage should also occur during the ripening of paraffin-coated cells. 
Paraffin 2 showed higher Cs peaks than paraffin 1. This trend, that coatings that adsorb high quantities of Cs do not have a sufficient anti-relaxation effect, was confirmed for other paraffin coatings fabricated in this experiment.

\begin{figure}[tb]
 \begin{center}
  \includegraphics[width=90mm]{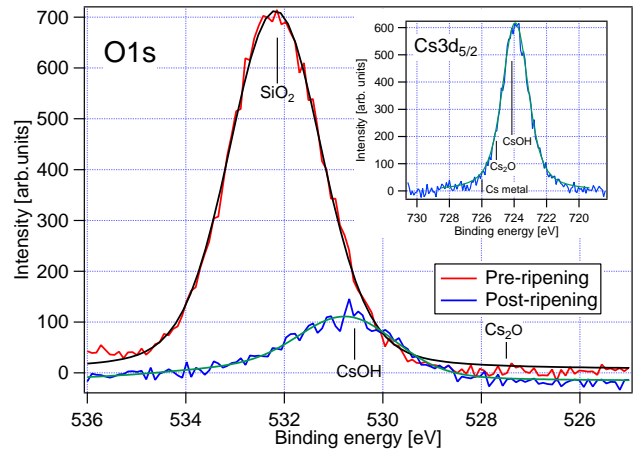}
 \end{center}
 
 \caption{XPS spectra of O 1s for paraffin 2 taken before and after the spin relaxation measurement. Fitting Voigt curves are also displayed. Binding energies for possible compounds  in the literature~\cite{NIST} are indicated. Inset: XPS spectrum of Cs 3d$_{5/2}$ for paraffin 2 taken after the spin relaxation measurement.}
 \label{Fig5}
\end{figure}

The O peak was observed after ripening, despite the other components contained in the glass substrate disappearing.
Figure~\ref{Fig5} shows detailed spectra for the O 1s peak measured for paraffin 2.
Before ripening, the binding energy of the O 1s electron corresponded to $\rm{SiO_2}$ in the glass substrate. After ripening, the peak was shifted to lower binding energies, which indicates that the observed O originated not from the glass substrate, but from O-Cs compounds.  The peak position of Cs 3d$_{5/2}$ shown in the inset of Fig.~\ref{Fig5}  also suggests that Cs was bound to O. Furthermore, we found a clear correlation between the Cs and O concentrations in all investigated paraffin coatings. Although the produced O-Cs compound is likely to be cesium hydroxide from the XPS spectra, we can not exclude other possibilities such as Cs suboxides, due to lack of XPS data on those oxides. We consider that the O, which was bound to Cs, was an impurity introduced during the coating characterization processes, including spin relaxation measurement and sample transfer in the glovebox. It has been difficult to suppress this oxygen contamination in our experiment.
O-Cs compounds may also present in sealed cells with anti-relaxation coatings. They were actually detected in non-coated sealed cells~\cite{Pat07}.

\subsection{Diamond-like carbon}
\label{DLC_1}

\begin{figure}[tb]
 \begin{center}
  \includegraphics[width=90mm]{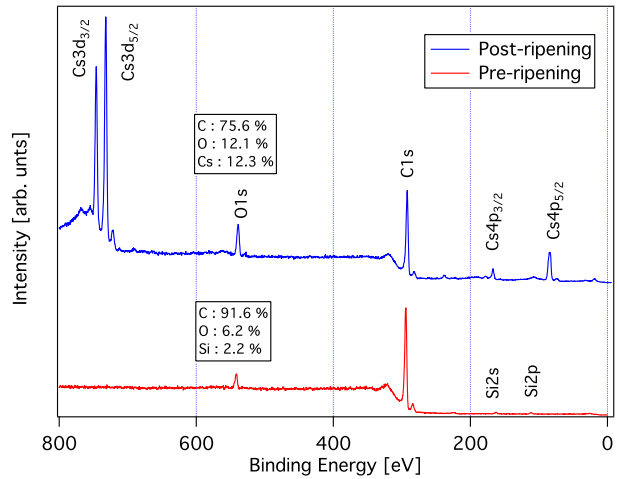}
 \end{center}
 \caption{XPS spectra of the DLC-coated slide taken before and after the spin relaxation measurement.}
 \label{Fig6}
\end{figure}

The DLC coatings we tried produced relaxation curves similar to those for non-coated glass slides, as shown in Fig.~\ref{Fig3}. We were unable to find any anti-relaxation effect for the DLC coatings. 
The XPS spectra in Fig.~\ref{Fig6} show that the DLC coating had a higher initial coverage than the paraffin 1 and paraffin 2 coatings shown in Fig.~\ref{Fig4}, but after ripening, more Cs atoms were observed with the DLC than with the paraffin. 
We found from XPS that only a small fraction of Cs atoms were lost from DLC coatings even when the slides were heated up to $400^\circ$C. This indicates that Cs atoms were strongly bound to the DLC coatings.

\section{Conclusions}
\label{conclusion}
From both spin relaxation measurement and surface analysis by XPS for each coating sample, we obtained the following results, which are useful to understand the operation of anti-relaxation-coated cells. According to XPS measurements, prepared paraffin coatings did not completely cover the glass substrates. The ripening process created a greater coverage of the paraffin coating on the substrates.
Coated slides with low effectiveness had less paraffin coverage after sample preparation and a higher level of Cs after the ripening process compared to the more effective coatings. During the coating characterization processes, O, which was probably contamination, was bound to Cs atoms adsorbed on the paraffin coatings. 
The DLC coatings did not show any antirelaxation effect. A relatively large amount of Cs was adsorbed on the DLC coatings. 

Although this research shows that high paraffin coverage is required for effective anti-spin relaxation performance, 
it is not clear that ``100\%'' paraffin coverage would directly lead to 1000, 10000, or even more bounces before atoms become depolarized. Further investigations using many coated slides fabricated with a highly reproducible coating method, possibly with additional surface analysis techniques, will be required to elucidate the physics of high antirelaxation performance.

\begin{acknowledgements}
The authors thank H. Usui and S. L. Bernasek for helpful discussions.
This work was supported by a Grant-in-Aid for Scientific Research (No. 23244082) from 
the Japan Society for the Promotion of Science, and also by the TUAT program ``Interdisciplinary Research Unit in Photon-Nano Science''.
\end{acknowledgements}




\end{document}